\documentclass[twocolumn,showpacs,preprintnumbers,superscriptaddress,amsmath,amssymb]{revtex4}
\usepackage{graphicx}% Include figure files
\usepackage{dcolumn}% Align table columns on decimal point
\usepackage{bm}% bold math

\begin{document}

\preprint{APS/123-QED}

\title{Hard X-ray Cu $2p$ Core-Level Photoemission of High-$T_c$ Cuprate Superconductors}

\author{M. Taguchi}

\affiliation{Soft X-ray Spectroscopy Lab, RIKEN SPring-8 Center, Sayo, Sayo, Hyogo 679-5148, Japan}

\author{M. Matsunami}

\affiliation{Soft X-ray Spectroscopy Lab, RIKEN SPring-8 Center, Sayo, Sayo, Hyogo 679-5148, Japan}

\author{A. Chainani}

\affiliation{Soft X-ray Spectroscopy Lab, RIKEN SPring-8 Center, Sayo, Sayo, Hyogo 679-5148, Japan}

\author{K. Horiba}

\affiliation{Soft X-ray Spectroscopy Lab, RIKEN SPring-8 Center, Sayo, Sayo, Hyogo 679-5148, Japan}

\author{Y. Takata}

\affiliation{Soft X-ray Spectroscopy Lab, RIKEN SPring-8 Center, Sayo, Sayo, Hyogo 679-5148, Japan}

\author{K. Yamamoto}

\affiliation{Soft X-ray Spectroscopy Lab, RIKEN SPring-8 Center, Sayo, Sayo, Hyogo 679-5148, Japan}

\author{R. Eguchi}

\affiliation{Soft X-ray Spectroscopy Lab, RIKEN SPring-8 Center, Sayo, Sayo, Hyogo 679-5148, Japan}

\author{M. Yabashi}

\affiliation{Coherent X-ray Optics Lab, RIKEN SPring-8 Center, Sayo, Sayo, Hyogo 679-5148, Japan}
\affiliation{JASRI/SPring-8, Sayo, Sayo, Hyogo 679-5198, Japan}

\author{K. Tamasaku}

\affiliation{Coherent X-ray Optics Lab, RIKEN SPring-8 Center, Sayo, Sayo, Hyogo 679-5148, Japan}

\author{Y. Nishino}

\affiliation{Coherent X-ray Optics Lab, RIKEN SPring-8 Center, Sayo, Sayo, Hyogo 679-5148, Japan}

\author{T. Nishio}
\affiliation{Institute of Materials Science, University of Tsukuba, Tsukuba, Ibaraki 305-8573, Japan}

\author{H. Uwe}
\affiliation{Institute of Materials Science, University of Tsukuba, Tsukuba, Ibaraki 305-8573, Japan}

\author{T. Mochiku}

\affiliation{National Institute for Materials Science, Tsukuba, Ibaraki 305-0047, Japan}

\author{K. Hirata}

\affiliation{National Institute for Materials Science, Tsukuba, Ibaraki 305-0047, Japan}

\author{J. Hori}

\affiliation{Department of Quantum Matter, ADSM, Hiroshima University, Higashi-Hiroshima, Hiroshima 739-8530, Japan}

\author{K. Ishii}

\affiliation{Department of Quantum Matter, ADSM, Hiroshima University, Higashi-Hiroshima, Hiroshima 739-8530, Japan}

\author{F. Nakamura}

\affiliation{Department of Quantum Matter, ADSM, Hiroshima University, Higashi-Hiroshima, Hiroshima 739-8530, Japan}

\author{T. Suzuki}

\affiliation{Department of Quantum Matter, ADSM, Hiroshima University, Higashi-Hiroshima, Hiroshima 739-8530, Japan}

\author{S. Shin}

\affiliation{Soft X-ray Spectroscopy Lab, RIKEN SPring-8 Center, Sayo, Sayo, Hyogo 679-5148, Japan}
\affiliation{Institute for Solid State Physics, University of Tokyo, Kashiwa, Chiba 277-8581, Japan}

\author{T. Ishikawa}

\affiliation{Coherent X-ray Optics Lab, RIKEN SPring-8 Center, Sayo, Sayo, Hyogo 679-5148, Japan}
\affiliation{JASRI/SPring-8, Sayo, Sayo, Hyogo 679-5198, Japan}

\date{\today} 

\begin{abstract}
We have performed a detailed study of Cu $2p$ core-level  spectra in single layer 
La$_{2-x}$Sr$_{x}$CuO$_{4}$, La doped Bi$_2$Sr$_{1.6}$La$_{0.4}$CuO$_{6+\delta}$ (Bi2201) and bilayer Bi$_2$Sr$_{2}$CaCu$_{2}$O$_{8+\delta}$ (Bi2212) high-temperature superconductors by using hard x-ray photoemission (HX-PES). We identify the Cu$^{2+}$ derived (i) the Zhang-Rice singlet (ZRS) feature, (ii) 
the $d^{n+1}\underline{L}$ (ligand screened) feature, (iii) the $d^{n}$ satellite feature, as well as the hole-doping derived high binding energy feature in the main peak.
 In Bi-based cuprates, intensities of the $d^{n}$ satellite features seem to be strongly enhanced compared to La$_{2-x}$Sr$_{x}$CuO$_{4}$. From x-ray photon energy dependent measurements, it is shown that the increased intensity in the satellite region is associated with Bi $4s$ core-level spectral intensity. 
The corrected $d^{n}$ satellite intensity is independent of the doping content or number of Cu-O layers. Our results suggest a correlation of the relative intensity of ZRS feature and hole-doping induced high binding energy spectral changes in the main peak with superconductivity.
  
\end{abstract}


\maketitle

\section{Introduction}

The electronic structure of the high-temperature (high-T$_c$) 
superconductors continues to remain a 
challenging topic in the physics of strongly correlated transition metal oxides.  Recent studies have 
elucidated important features involving electron-phonon coupling and charge-order \cite{lan01},
as well as two dimensionality of the magnetic excitations in the high-T$_c$ cuprates \cite{hin04}. The Zhang-Rice singlet (ZRS),  is a basic starting point for understanding the electronic structure of the cuprates. The ZRS was originally postulated \cite{zha88} 
for a doped hole distributed over the oxygen sites in a CuO$_4$ plaquette, which gets bound to the Cu$^{2+}$ ($3d^{9}$) state and forms a singlet due to Cu-O hybridization. It is the lowest energy electron removal state. Since a photoemission process also 
removes an electron (adds a hole), it was nicely shown that the ZRS can
be probed by valence band photoemission spectroscopy even for
insulating copper oxides \cite{wel95,pot97}. 
Polarization dependent \cite{gol97} and spin-resolved \cite{bro01} 
valence band photoemission have established
the ZRS feature as the lowest energy state in insulating and superconducting copper oxides.
It was also predicted that 
Cu $2p$ core-level photoemission spectra should exhibit a low binding energy feature derived from the ZRS, even for the parent insulating La$_{2}$CuO$_{4}$ \cite{Vee93}.
While the theoretical prediction was made very early, only recent studies with 
hard x-ray photoemission spectroscopy (HX-PES) could identify the ZRS feature in 
the Cu $2p$ core-level spectra of the parent La$_{2}$CuO$_{4}$ \cite{tag05PRL}.

HX-PES has developed significantly in the last few years \cite{pro05} 
mainly due to its ability to overcome the high-surface sensitivity of
conventional photoemission spectroscopy. HX-PES results in much higher 
kinetic energies of emitted electrons and hence longer mean free paths ( typically 50 - 150 $\AA$ ),
naturally leading to bulk-sensitive measurements \cite{kob03,tak04,dal04}. 
Its application to strongly correlated systems has led to several significant results, 
making it an fundamental technique to reveal the true bulk electronic structure 
\cite{cha04,hor04,kam04,tag05PRB,tag05PRL,tan06,pan06}. 
In the last few years, we have reported bulk sensitive HX-PES with $h\nu$=5.95 keV of 
transition metal (TM) $2p$ core-level for various TM compounds 
(pure and Cr doped V$_2$O$_3$ \cite{kam04,tag05PRB}, hole-doped LaMnO$_3$ (LSMO, LBMO) \cite{hor04,tan06} and high-Tc cuprates \cite{tag05PRB,tag05PRL}). Unlike conventional soft x-ray (SX) PES, TM $2p$ core-level HX-PES spectra have shown additional well-screened satellites with significantly large intensity in addition to $2p^53d^n$ and $2p^53d^{n+1}\underline{L}$ features, where $\underline{L}$ represents a ligand hole. These satellites were located at an energy position few eV lower in binding energy than the $2p_{3/2}$ and $2p_{1/2}$ main line, respectively. 

In the case of pure and Cr doped V$_2$O$_3$, clear changes in V $2p$ HX-PES were observed across the metal-insulator (Mott-Hubbard) transition \cite{kam04,tag05PRB}. The structure and position of the main peak and the high and low binding energy satellite structures were explained well by the cluster model including intra-atomic multiplet structure and a metallic screening channel derived from coherent states at the Fermi energy. The low binding energy satellite structure was attributed to the screening from the coherent state at the Fermi level. This has been independently confirmed by direct measurements of the 
valence band coherent feature and core-level spectra of  V$_2$O$_3$ \cite{pan06}. A similar interpretation based on a Mott-Hubbard transition has been used to explain the Ru $3d$ core-level spectra for a series of Ru oxides \cite{kim04}.

For hole-doped LaMnO$_3$, a similar well-screened low binding energy feature in Mn $2p$ spectra was observed \cite{hor04}. The feature showed a noticeable increase with decreasing temperature on entering the ferromagnetic metal phase, indicating that the origin of this feature is strongly related to the ferromagnetic metal phase \cite{hor04,tan06}. 

New features at low binding energy are also seen in high-T$_c$ cuprates \cite{tag05PRL}. But its spectral shape is quite different in electron-doped Nd$_{1.85}$Ce$_{0.15}$CuO$_{4}$ (NCCO) and hole-doped La$_{1.85}$Sr$_{0.15}$CuO$_{4}$ (LSCO) systems. It was clearly shown that LSCO and NCCO exhibit new bulk character electronic states, which are strongly suppressed in the surface.

A recent theoretical study \cite{vee06} offers an alternative interpretation for well-screened features in the Cu $2p$, Mn $2p$, and Ru $3d$ core-level spectra based on a non-local screening scenario with doping, magnetic and orbital ordering for the cuprates, manganites and ruthenates, respectively. We believe in so far as the additional screening channel is required to explain the low binding energy feature, our picture is not different from Ref.~\cite{vee06}. 
In particular, we believe the non-local screening channel in the multi-site cluster model \cite{vee06} would develop into quasiparticle states at the Fermi level ($E_{F}$) as obtained in dynamical mean-field theory calculations \cite{kim04, cor06}. In the spirit of the non-local screening channel or screening from quasiparticle states at $E_{F}$, we have used a single-site cluster model \cite{tag05PRL, tag05PRB}, with the screening from the states at $E_{F}$ but also including intra-site multiplets which are necessary for the observed structure in the core-level spectra.

In a previous study, we have presented results of Cu $2p$ HX-PES for Bi$_2$Sr$_{2}$CaCu$_{2}$O$_{8+\delta}$ (Bi2212) \cite{tag05PRB}. We found that the spectral weight in the $"high"$ binding energy satellite increases considerably in HX-PES compared with SX-PES results. This enhancement of the satellite intensity in Bi2212 is quite different as opposed to the other TM compounds mentioned above. V$_2$O$_3$, LSMO and NCCO have shown huge changes at the $lower$ binding energy to the main peak, but Bi2212 shows a change in the $higher$ binding energy satellite. This fact would naively indicate an increase of $3d^9$ weight in the bulk since the satellite is generally assigned to the $2p^53d^9$ state. The fitted parameter values indicated that hybridization strength was reduced in the bulk sensitive HX-PES as against the surface sensitive SX-PES.  As previously mentioned \cite{tag05PRB}, a decrease of the hybridization strength in the bulk is surprising. In general, the different atomic environment and reduced co-ordination, often conspire to reduce hybridizations and screenings at the surface. 

In order to shed new light on the above-mentioned problem, we extend our previous work to study the Cu-O layer-dependence, doping dependence and photon energy dependence by using HX-PES. We report the Cu $2p$ core-level  spectra in single layer 
La$_{2-x}$Sr$_{x}$CuO$_{4}$ (LSCO, $x$ = 0.0, 0.145 and 0.26)
, La doped Bi$_2$Sr$_{1.6}$La$_{0.4}$CuO$_{6+\delta}$ (Bi2201) 
 and bilayer Bi2212. 
Our main purpose is to resolve the puzzling observation of increase in the spectral intensity of the
$3d^9$ satellite in Bi2212 by comparative measurements. 
From x-ray photon energy dependent measurements, 
the results show that the increased intensity in the satellite region is associated with Bi $4s$ core-level spectral intensity, which is usually neglected in SX-PES measurements.
The corrected $d^{n}$ satellite intensity is independent of the doping content or number of Cu-O layers. The results suggest a correlation of the relative intensities of the ZRS feature and the hole-doping induced high binding energy spectral changes in the main peak with superconductivity.   

The present paper is organized as follows. The experimental details are briefly described in Sec. II. The results and discussion on Cu-O layer dependence, doping dependence, and photon energy dependence are given in Sec. III. Conclusions are drawn in Sec. IV.

\section{Experiment}

HX-PES measurements were performed 
in a vacuum of 1 $\times$ 10$^{-10}$ Torr
at undulator beam line BL29XU, SPring-8 \cite{Tam01} using a Scienta R4000-10KV electron analyzer. The instrumentation details are described in Ref.~\cite{tak05, ish05}. The energy width of incident x-ray was 70 meV, and the total energy resolution, $\Delta$E was set to $\sim$ 0.4 eV. All spectra were corrected by subtracting a Shirley-type background. 
  Sample temperature was controlled to $\pm 2 $K during measurements. Single crystal Bi2212 and Bi2201 were peeled with a scotch tape and measured at 30 K. The LSCO samples were measured for surfaces  fractured $in$-$situ$. The $x$ = 0.0 composition was measured at room temperature to avoid charging while the metallic compositions ($x$ = 0.145 and 0.26) were measured at 30 K. A reference measurement on a metallic Bi-based non-cuprate oxides Ba$_{0.55}$K$_{0.45}$BiO$_3$ (BKBO) was measured at 30 K for an $in$-$situ$ fractured surface. The Fermi level of gold was measured to calibrate the energy scale.

\begin{figure}
\includegraphics[scale=.55]{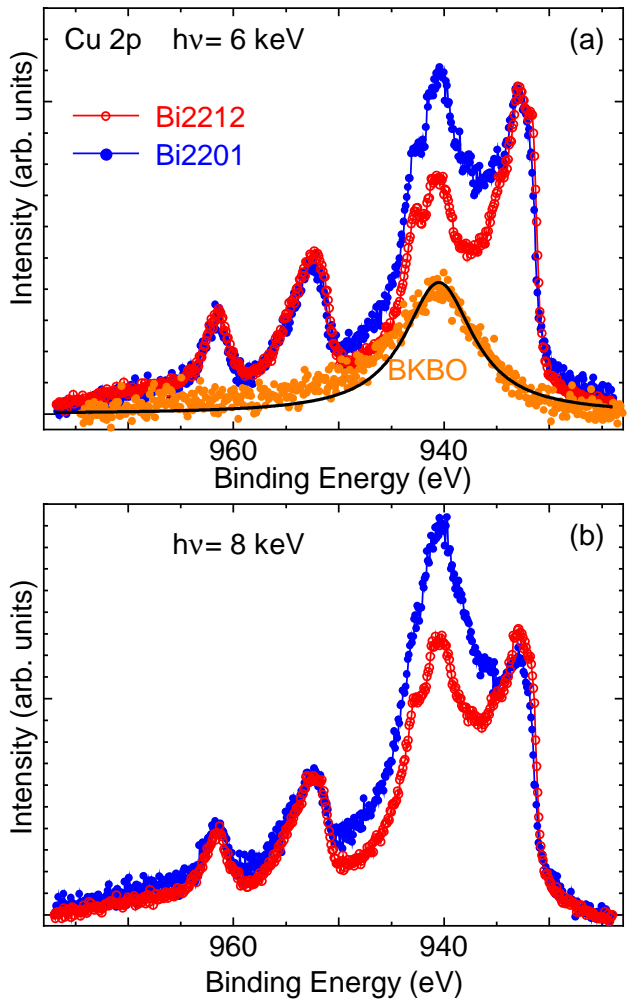}
\caption{\label{fig1}   %FIG.1
(Color online) (a) Cu $2p$ core-level HX-PES spectra of Bi2212, Bi2201 and Bi $4s$ HX-PES of BKBO (filled circles) with a photon energy $h\nu=5.95$ keV. The line is fit to the Bi $4s$ HX-PES data as explained in the text. (b) Cu $2p$ core-level HX-PES spectra of Bi2212 and Bi2201 with a photon energy $h\nu=7.93$ keV. All Cu $2p$ spectra were corrected by subtracting a Shirley-type background.}
%\vspace{-0.25in}
\end{figure}

\section{Results and Discussion}
\subsection{HX-PES in Bi2201 and Bi2212}
In Fig.~1(a) we present the Cu $2p$ core-level HX-PES spectra of Bi2212 and Bi2201 with photon energy $h\nu=5.95$ keV. The spectrum of Bi2212 is consistent with our previous report \cite{tag05PRB}. It consists of two prominent sets of structures and the main peaks at $\sim$ 934 and 953 eV, separated by approximately 19 eV, originating from the spin-orbit splitting of the $2p$ core-level, $2p_{3/2}$ and $2p_{1/2}$. Each prominent structure itself consists of two peaks, main peak ($2p^53d^{10}\underline{L}$ state) and satellite ($2p^53d^9$ state) nearly 7 eV above two main peaks. The spectral weight in the high binding energy satellite of $2p_{3/2}$ at 940 eV is significantly higher in HX-PES as compared with the previous SX-PES \cite{tag05PRB,Vee93,Koi02} and basically no change is seen over the $2p_{1/2}$ satellite energy range. 
In the previous paper \cite{tag05PRB}, we had carried out the cluster model calculation of Cu $2p$ HX-PES for Bi2212. To obtain the significant increase in the satellite intensity at 940 eV, a reduced hybridization strength was needed (by a factor of $\sim 0.8$) compared to that in the SX-PES, implying a decrease of the hybridization strength in the bulk. Since a smaller hybridization is expected on the surface compared to that in the bulk, this fact is quite unusual. 

From a comparison between Bi2212 and Bi2201, we found that the satellite intensity of Bi2201 is higher than that of Bi2212, as shown in Fig.~1(a). Even more surprisingly, its relative weight is almost comparable to the main line. This huge enhancement cannot be explained by the reduced hybridization in the bulk. 
Thus, we are led to another possibility for elucidating our observations. One possibility would be the photo-ionization cross section effect of Bi $4s$ core-level, because the binding energy of Bi $4s$ core-level is very close to the $2p^53d^9$ satellite energy of $2p_{3/2}$.

The Bi $4s$ contribution in Cu $2p$ core-level spectra of the Bi-based cuprates is generally neglected. This is a good approximation, because the small Bi $4s$ photo-ionization cross section leads to negligible intensity as far as the SX-PES is concerned. In fact, the cross section ratio $\sigma_{Bi}(4s)/\sigma_{Cu}(2p)$ in soft x-ray region is at most $\sim0.09$ \cite{yeh85}.  The increase of the x-ray photon energy will give an increase of the cross section ratio $\sigma_{Bi}(4s)/\sigma_{Cu}(2p)$. The $\sigma_{\rm Bi}(4s)/\sigma_{\rm Cu}(2p)$ ratio is very large and is approximately 1 in hard x-ray region. This means that Bi $4s$ contribution can by no means be neglected. Moreover, the composition ratio of Bi:Cu is 2:2  for Bi2212. Thus, we expect comparable contributions from the Bi $4s$ core-level respect to the Cu $2p$ spectra and was overlooked by us in the previous study. 

The considerable enhancement of the satellite intensity of Bi2201 with respect to Bi2212 can be similarly understood by comparing the composition ratio between Bi2201 and Bi2212. The composition ratio of Bi:Cu is 2:1 for Bi2201 and consequently, the relative contribution from the Bi $4s$ is enhanced by a factor of 2 with respect to Bi2212.

 Since, with further increase of the x-ray photon energy, the photo-ionization cross section ratio $\sigma_{\rm Bi}(4s)/\sigma_{\rm Cu}(2p)$ increases further, we expect the satellite intensity to be increased
with higher value of $h\nu$  compared to $h\nu=5.95$ keV photon energy. To confirm this, we performed HX-PES measurements by using 7.93 keV x-ray photon energy. 
Figure~1(b) shows the Cu $2p$ core-level HX-PES spectra for Bi2212 and Bi2201 using photon energy $h\nu=7.93$ keV. Again, the intensities at 940 eV for both Bi2212 and Bi2201 are clearly enhanced and they are much stronger than those with $h\nu=5.95$ keV, as we expected. In particular, for Bi2201, the intensity in the satellite region exceeds that of the main line and 
the satellite peak dominates over the main peak, confirming the Bi 4s contribution.

To correct for the Bi $4s$ contribution in the Cu $2p$ HX-PES spectra of the 
B-based cuprates, we measured the Bi $4s$ core-level data taken from a single crystal BKBO in Fig.~1(a). Since the measured signal-to-noise (S/N) ratio of BKBO spectrum is not sufficient for the present correction procedure, the Bi $4s$ spectrum of BKBO was fitted with a Lorenztian (HWHM=4 eV). 
The result of the fit is illustrated by a line in Fig.~1(a). 
This Lorenztian line shape was used to fit the Cu $2p$ HX-PES spectra for Bi2212 and Bi2201 with $h\nu=5.95$ keV. 
Then, we obtained the Cu $2p$ line shape by subtracting the best fit to the Bi $4s$ component, the only free parameter being the intensity ratio. 
Figure~2(a) shows the corrected Cu $2p_{3/2}$ HX-PES spectra of Bi2212 and Bi2201 compared with SX-PES data. Thus,
satisfactory corrections were obtained by subtracting the Bi $4s$ signal.

The obtained ratio of the integrated intensities of the 
subtracted Bi $4s$ Lorenztian contribution between Bi2212 and Bi2201 is 1:2, which is consistent with the relative Bi contribution between two materials. 
The resulting Cu $2p$ HX-PES for Bi2212 and Bi2201 are quite similar to those of SX-PES for Bi2212.

\begin{figure}
\includegraphics[scale=.55]{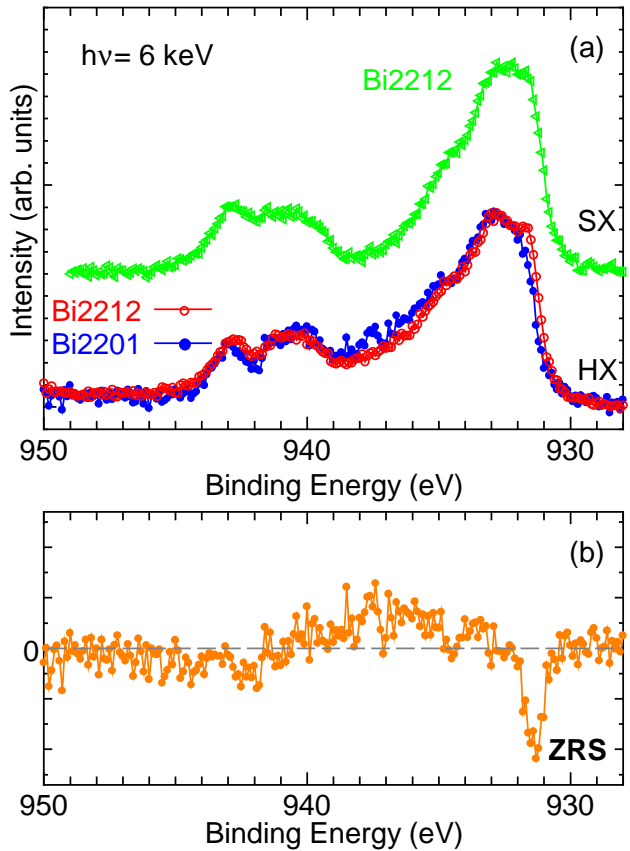}
\caption{\label{fig2}     %FIG.2
(Color online) (a) Comparison between experimental Cu $2p_{3/2}$ SX-PES from ref.~\cite{tag05PRB} and Cu $2p_{3/2}$ HX-PES after subtracting Bi $4s$ contribution. (b) Difference between Cu $2p$ HX-PES spectra of Bi2201 and that of Bi2212. The dip feature at 931 eV is the ZRS feature. The broad low intensity feature at higher binding energy is due to the poor S/N ratio in the Bi2201 spectrum. }
%\vspace{-0.25in}
\end{figure}

\begin{figure}
\includegraphics[scale=.55]{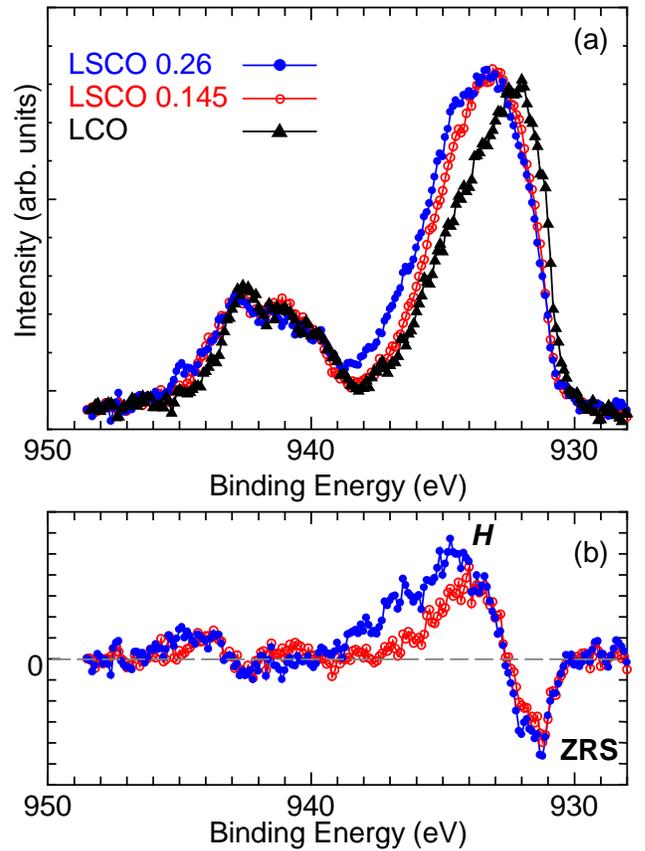}
\caption{\label{fig3}     %FIG.3
(Color online) (a) Cu $2p_{3/2}$ HX-PES spectra of La$_{2-x}$Sr$_x$CuO$_4$ for several doping with a photon energy $h\nu=5.95$ keV. (b) Difference spectra between several Cu $2p_{3/2}$ HX-PES spectra and reference spectrum of LCO.}
%\vspace{-0.25in}
\end{figure}

Having subtracted the Bi $4s$ contribution, we now concentrate on the detailed structure in the main peak of Cu $2p_{3/2}$. The Bi2212 spectrum [Fig.~2(a)] shows clear changes compared to Bi2201 and the additional spectral weight is seen at the lower binding energy (nearly 931 eV). This is particularly evidenced well by the difference spectra between Bi2201 and Bi2212 in Fig.~2(b). Using the multi-site cluster model \cite{Vee93,vee94,oka97}, it was shown that the Cu $2p$ spectrum has a low binding energy ZRS feature due to the non-local screening. Thus, this feature is assigned to the ZRS feature, indicating that the ZRS feature is enhanced in the bilayer Bi2212 compared to the single layer Bi2201. 

\subsection{HX-PES in LSCO}

In order to better understand the character of the main line, we resort to a comparison with the LSCO Cu $2p$ spectra obtained with HX-PES. Cu $2p_{3/2}$ HX-PES spectra of LSCO as a function of doping are shown in Fig.~3(a). The spectra of LSCO with $x$ = 0.0  and 0.145 are the same as those of our previous report \cite{tag05PRL}. We observed a systematic change in spectral shape on hole-doping.  To see these spectral features clearly we plot the difference spectra with respect to the LCO reference spectrum in Fig.~3(b). 
The observed spectral changes in our data are consistent with previous works on the YBa$_2$Cu$_3$O$_{7-\delta}$ \cite{gou88, fla89}. 
The shoulder at lower energy to the main peak $2p^53d^{10}\underline{L}$ in LSCO can again be similarly attributed to the ZRS feature, consistent with earlier work \cite{Vee93}.
As discussed in our earlier study, the ZRS peak in 
$x$ = 0.0 occurs at the lower binding energy to the $2p^53d^{10}\underline{L}$ state. The spectral intensity in the ZRS 
feature  reduces on increasing $x$ from $x$ = 0.0 to 0.145, but does not change for $x$ = 0.26.

Note that Fig.~3(a) shows another feature, $ie.$, a broad feature centered at 934.5 eV (feature $H$) which is enhanced as the hole-doping increases. In contrast to the ZRS feature, the feature $H$ grows with increasing $x$ up to $x$ = 0.26.

We suggest two possible explanations for this broad feature $H$. In LSCO, since we dope holes due to Sr content $x$, it can lead to the Cu$^{3+}$ character state which grows with doping. 
Another possible explanation is the non-local screening effects. Very recently, Okada and Kotani developed the multi-site cluster model calculation for the doping dependence \cite{oka05} and a similar enhancement as observed for the high binding energy feature was already apparent in their calculation. 
However the relation of non-local screening effects with Cu$^{3+}$ configurations is not clear. 
We cannot conclude from our experiments alone if the feature $H$ reflects the non-local screening effect or simply an increase in  Cu$^{3+}$ configurations with hole-doping, or a coupling of both effects.

\subsection{Comparison of Bi2201, Bi2212 and LSCO}

A comparison of LSCO spectra indicates that the ZRS feature intensity between $x$ = 0.145 (T$_c$=36 K) and 0.26 (non superconducting) does not change much but the intensity of the feature $H$ beyond optimal doping increases. In terms of its superconducting $T_c$, it 
then suggests that an increase in the feature $H$ content is primarily responsible for the reduction in $T_c$.
On the other hand, for Bi2201 ($T_c$=28 K) and Bi2212 ($T_c$=90 K), the high binding energy feature does not seem to change while the ZRS feature shows a small but definite increase in intensity for Bi2212. This suggests that the retention or increase of the ZRS feature is more important for higher Tc in the hole doped cuprates.

\section{Summary}

In conclusion, the aim of the present paper has been to understand the origin of the experimentally observed satellite at 940 eV in Cu $2p$ HX-PES of Bi based high-T$_c$ superconductors and in particular their evolution with x-ray photon energy and the composition change. 
We investigated the photon energy dependence of the Cu $2p$ core-level photoemission spectra for Bi2212 and Bi2201 by using hard x-ray photon energy. We exploited the sensitivity of the relative photo-ionization cross section to clarify the Bi $4s$ contribution from the measured Cu $2p$ HX-PES spectra. The results from Bi based cuprates can be interpreted satisfactorily only by including the Bi $4s$ core-level contribution. 
Thus the observed enhancement of the satellite is then due to the increase of the photo-ionization cross section of Bi $4s$ core-level rather than the reduction of the hybridization strength in the bulk. 
The corrected spectra lead to a systematic picture on the basis of the doping dependence and Cu-O layers dependence of ZRS feature and hole-doping induced high binding energy spectral changes in the main Cu $2p$ core-level line.

\section{Acknowledgment}

The HX-PES experiments reported here have benefited tremendously from the efforts of Dr. D. Miwa of the coherent x-ray optics laboratory RIKEN/SPring-8, Japan and we dedicate this work to him.

\end{document}